\documentstyle[12pt]{article}
\renewcommand{\baselinestretch}{1.5}

\normalsize
\def\etal{{\it et\thinspace al.}\ }
\begin{document}
\title{Large-scale Breit-Pauli R-matrix calculations for transition 
probabilities of Fe V}
\author{Sultana N. Nahar and Anil K. Pradhan\\
Department of Astronomy, The Ohio State University\\
Columbus, Ohio 43210, U.S.A.}
\maketitle
\begin{abstract}
\end{abstract}
 
Ab initio theoretical calculations are reported for the electric (E1) dipole 
allowed and intercombination fine structure transitions in Fe V 
using the Breit-Pauli R-matrix (BPRM) method. We obtain 3865 bound fine 
structure levels of Fe V and $1.46\times 10^6$ oscillator strengths, 
Einstein A-coefficients and line strengths. In addition to the 
relativistic effects, the 
intermediate coupling calculations include extensive electron 
correlation effects that represent the complex configuration interaction 
(CI). For bound-bound transitions the BPRM method, based on atomic 
collision theory, entails the computation of the CI wavefunctions 
of the atomic system as an (electron~+~target ion) complex. The target 
ion Fe~VI is represented by an eigenfunction expansion of 19 fine 
structure levels dominated by the spectroscopic configuration $3d^3$, 
and a number of correlation
configurations. Fe~V bound levels are obtained with angular and 
spin symmetries $SL\pi$ and $J\pi$ of the (e~+~Fe~VI) system such that 
$2S+1$ = 5,3,1, $L \leq$ 10, $J \leq 8$. The bound levels are obtained as
solutions of the Breit-Pauli (e~+~ion) Hamiltonian for each $J\pi$, 
and are designated according to the `collision' channel quantum numbers.
A major task has been the identification of these large number of bound
fine structure levels in terms of standard spectroscopic designations. 
A new scheme, based on the analysis of quantum defects and channel 
wavefunctions, has been developed. The identification scheme aims 
particularly to determine the completeness of the results in terms of 
all possible bound levels with $n \leq 10, l \leq n-1$, for applications 
to analysis of experimental measurements and plasma modeling. 
Sample results are presented and the accuracy of the results is 
discussed. A comparison of the dipole length and velocity oscillator 
strengths is presented, indicating an uncertainty of
10-20\% for most transitions.

\section{Introduction}

Transition probabilities of heavy elements, particularly the iron group, 
are of great importance in astrophysical and laboratory sources. Fuhr 
et al. [1] have compiled data from a number of available sources.
However, the accuracy and the extent of these data is largely inadequate
for many general applications such as the calculation of local thermodynamic
equilibrium (LTE) stellar opacities [2,3], and radiative
levitation and accelerations of heavy elements [4]. Among the particular 
applications including Fe~V as a prominent spectral constituent are the
the non-LTE models of Fe V spectra in hot stars [5], and the observed
extreme ultraviolet Fe~V emission from young white dwarfs [6]. For
example, currently
available data for Fe~V fails to account for the observed opacity of
iron in the XUV region where observations of newly formed hot and 
young white dwarfs clearly show Fe~V lines [6].
In all of these applications it is highly desirable to have as
complete a dataset of radiative transition probabilities as possible.
While the twin problems of completeness and accuracy pose a
challenge to the theoretical methods, they are of interest 
not only in various applications but may also be of use in the analysis
of experimental measurements of observed energy levels of complex atomic
systems from the iron group.

The Opacity Project (OP) [7,2] and the Iron Project (IP) [8] laid the 
foundation for large-scale theoretical calculations using ab intio methods.
The R-matrix method [9], based on atomic collision theory
techniques and adapted for the OP [10] and the IP [8], 
has proven to be very efficient for these calculations. Whereas the OP
calculations were all in the LS coupling approximation, with no
relativistic effects included, the subsequent IP work is in intermediate
coupling using the Breit-Pauli extension of the R-matrix method [8].
While most the IP work has concentrated on collisional
calculations, recent works have extended the BPRM method to radiative
bound-bound and bound-free calculations for transition probabilities 
[11], photoionization [12], and (electro-ion) recombination [13]. The first 
comprehensive BPRM calculation of fine structure transition 
probabilities was carried out for the highly charged ions Fe~XXIV and 
Fe~XXV [11] that are 
of special interest in X-ray astronomy. Very good agreement was found
with existing results available for a limited number of transitions but
using very accurate theoretical methods including relativistic and QED
effects [14,15], thus establishing the
achievable accuracy for the BPRM calculations. However those He-like
and Li-like atomic systems are relatively simple, and the electron
correlation effects relatively weak, compared to the low ionization
stages of iron group elements. The present work attempts to enlarge the
scope of the possible BPRM calculations to include the iron group
elements, as well as to solve some outstanding problems related to level
identifications in {\it ab initio} theoretical calculations using collision
theory methods.

Unlike atomic structure calculations, where the electronic configurations 
are pre-specified and the levels identified, the bound levels
calculated by collision theory methods adopted in the OP and the IP need
to be identified since only the channel quantum numbers are known for the
bound states corresponding to the (e + ion) Hamiltonian of a given total
angular and spin symmetry $SL\pi$ or $J\pi$. The precise correspondence between
the channels of the collision complex, and the bound levels, 
must therefore be determined. The problem is non-trivial for complex
atoms and ions with many highly mixed levels due to configuration
interaction. In the OP work, carried
out in LS coupling, this problem was solved by an analysis based on
quantum defects and the numerical components of wavefunctions
in the region outside the R-matrix boundary (that envelops the target
ion orbitals). The present work extends that treatment to the analysis
of fine structure levels computed in intermediate coupling. In
addition, considerable effort is devoted to the determination of the
completeness of the set of computed bound levels; comparison with the
expected levels derived from all possible combination of angular and
spin quantum numbers reveals the missing levels. The general procedure
could be applied to spectroscopic measurements and the 
analysis of observed levels of a given atomic system by comparison with
the theoretical predictions.

\section{Theory}

The general theory for the calculation of bound states in the close
coupling (CC) approximation of atomic collision theory, using the
R-matrix method, is described by Burke and Seaton [16] and Seaton [17]. 
The application to the Opacity Project work is described by Seaton [7], 
Berrington \etal [10], and Seaton \etal [2]. The relativistic extensions 
of the R-matrix method in the Breit-Pauli approximation are discussed 
by Scott and Taylor [18], and the computational details by Berrington, 
Eissner, and Norrington [19]. The application to the Iron Project
work is outlined in Hummer \etal [8].

In the present work we describe the salient features of the theory and
computations as they pertain to large-scale BPRM calculations for complex
atomic systems. Identification of fine structure energy levels is 
discussed in detail.

Following standard collision theory nomenclature, we refer to the (e +
ion) complex in terms of the 'target' ion, with N bound electrons,
and a 'free' electron that may be either bound or continuum. 
The total energy of the system is either negative or positive; negative
eigenvalues of the (N + 1)-electron Hamiltonian correspond to 
bound states of the (e + ion) system. In the coupled channel or close 
coupling (CC) approximation the wavefunction expansion, $\Psi(E)$, for 
a total spin and angular symmetry  $SL\pi$ or $J\pi$, of the (N+1) 
electron system is represented in terms of the target ion states as:

\begin{equation}
\Psi_E(e+ion) = A \sum_{i} \chi_{i}(ion)\theta_{i} + \sum_{j} c_{j} \Phi_{j},
\end{equation}

\noindent
where $\chi_{i}$ is the target ion wave function in a specific state
$S_iL_i\pi_i$ or level $J_i\pi_i$, and $\theta_{i}$ is the wave function
for the (N+1)th electron in a channel labeled as
$S_iL_i(J_i)\pi_i \ k_{i}^{2}\ell_i(SL\pi) \ [J\pi]$; $k_i^2(=\epsilon_i)$ 
is the incident kinetic energy. In the second sum the $\Phi_j$'s are
correlation
wavefunctions of the (N+1) electron system that (a) compensate for the
orthogonality conditions between the continuum and the bound orbitals,
and (b) represent additional short-range correlation that is often of
crucial importance in scattering and radiative CC calculations for each
$SL\pi$. 

The functions $\Psi(E)$ are given by the R-matrix method in an inner
region $r \leq a$. These are bounded at the origin and contain radial
functions that satisfy a logarithmic boundary condition at $r = a$ [20].
In the outer region $r > a$ the inner region functions 
are matched to a set of linearly independent functions that correspond 
to all possible (e~+~ion) channels of a given symmetry $SL\pi$ or 
$J\pi$.  The outer region wavefunctions are computed for all channels, 
$(C_tS_tL_t\pi_t)\epsilon l$, where $C_t$ is the target configuration, 
and used to determine the individual channel contributions (called 
``channel weights").

In the relativistic BPRM calculations the set of ${SL\pi}$
are recoupled to obtain (e + ion) levels  with total $J\pi$, followed by
diagonalisation of the (N+1)-electron Hamiltonian, 
\begin{equation}
H^{BP}_{N+1}\mit\Psi = E\mit\Psi.
\end{equation}
The BP Hamiltonian is
\begin{equation}
H_{N+1}^{\rm BP}=H_{N+1}+H_{N+1}^{\rm mass} + H_{N+1}^{\rm Dar}
+ H_{N+1}^{\rm so},
\end{equation}
where $H_{N+1}$ is the nonrelativistic Hamiltonian,
\begin{equation}
H_{N+1} = \sum_{i=1}\sp{N+1}\left\{-\nabla_i\sp 2 - \frac{2Z}{r_i}
        + \sum_{j>i}\sp{N+1} \frac{2}{r_{ij}}\right\},
\end{equation}
and the additional terms are the one-body terms, the mass correction term, 
the Darwin term and the spin-orbit term respectively. Spin-orbit 
interaction, $H^{so}_{N+1}$, splits the LS terms into fine-structure 
levels labeled by $J\pi$, where $J$ is the total angular momentum.
Other terms of the Breit-interaction [22],
\begin{equation}
H^B = \sum_{i>j}[g_{ij}({\rm so}+{\rm so}')+g_{ij}({\rm ss}')],
\end{equation}
representing the two-body spin-spin and the spin-other-orbit 
interactions are not included. 

The positive and negative energy states (Eq. 1) define continuum or
bound (e~+~ion) states,

\begin{equation}
 \begin{array}{l} E = k^2 > 0  \longrightarrow
continuum~(scattering)~channel \\  E = - \frac{z^2}{\nu^2} < 0
\longrightarrow bound~state, \end{array}
\end{equation}
where $\nu$ is the effective quantum number relative to the core level.
If $E <$ 0 then all continuum channels are `closed' and the solutions
represent bound states. 
Determination of the quantum defect ($\mu(\ell))$, defined as  
$\nu_i = n - \mu(\ell)$ where $\nu_i$ is relative to the core level 
$S_iL_i\pi_i$, is helpful in establishing the $\ell$-value associated 
with a given channel (level).

At E $<$ 0 a scattering channel may represent a bound state at the proper
eigenvalue of the Hamiltonian (Eq. 2). 
A large number of channels are considered for the radiative processes of
Fe V. Each SL$\pi$ or J$\pi$ symmetry is treated independently and 
corresponds to a large number of
channels. Therefore, the overall configuration interaction included in
the total (e + ion) wavefunction expansion is quite extensive. This is
the main advantage of the CC method in representing electron correlation
accurately.

\subsection{Level identification and coupling schemes}

The BPRM calculations in intermediate coupling employ the pair-coupling
representation

\begin{equation}
 \begin{array}{l} S_i + L_i \longrightarrow J_i \\
 J_i + \ell \longrightarrow K \\
 K + {\rm s} \longrightarrow J \end{array},
\end{equation}
where the `i' refers to the target ion level and $\ell, s$ are the
orbital angular momemtum (partial wave) and spin of the additional electron.
According to designations of a collision complex, a channel is fully
specified by the quantum numbers
\begin{equation}
(S_iL_i \ J_i)\pi_i \ \epsilon_i \ \ell_i \ K~s \ [J\pi]
\end{equation}

 The main problem in identification of the fine structure levels stems 
from the fact that 
the bound levels are initially given only as eigenvalues of the
(e~+~ion) Hamiltonian of a given symmetry $J\pi$. Each level therefore
needs to be associated with the quantum numbers characterizing a 
given collision channel. Subsequently, three main parameters are to be 
determined: (i) the parent or the target ion level, (ii) the orbital,
effective and principal quantum numbers $(l, \nu, \ n)$ of the (N+1)th
electron, and (iii) the 
symmetry, $SL\pi$. The task is relatively straightforward
for simple few-electron atomic systems. For example, in a recent work
Nahar and Pradhan [11] have calculated a large number of 
transition probabilities for Li-like Fe~XXIV and He-like Fe~XXV, where
the problem of level identification is trivial, compared to the present
work, since the bound levels are well separated in energy and in $\nu$. 
However when a number of mixed bound levels fall within a given interval
$(\nu, \ \nu+1)$, for the same $J\pi$, the quantum numbers and the 
magnitude of the components in all associated channels must be analysed.
A scheme for identification of levels is developed (discussed later)
that rests mainly on 
an analysis of quantum defects of the bound levels and their orbital 
angular momenta, and the percentage of the total wavefunction in all 
channels of a given $J\pi$.

Following level identification, further work is needed to enable a 
direct correspondence with standard spectroscopic designations that 
follow different coupling schemes, such as between $LS$ and $JJ$, 
appropriate for atomic structure calculations as, for example, in the 
NIST tables of observed energy levels [1]. The correspondence provides
the check for completeness of calculated set of levels or the levels
missing. The level identification procedure involves considerable 
manipulation of the bound level data and, although it has been encoded 
for general applications, still requires analysis and interpretation of 
problem cases of highly mixed levels that are difficult to identify.

\subsection{Oscillator strengths and transition probabilities}

The oscillator strength (or photoionization cross section) is
proportional to the generalized line strength defined,
in either length form or velocity form, by the equations
\begin{equation}
S_{\rm L}=
 \left|\left\langle{\mit\Psi}_f
 \vert\sum_{j=1}^{N+1} r_j\vert
 {\mit\Psi}_i\right\rangle\right|^2 \label{eq:SLe}
\end{equation}
and
\begin{equation}
S_{\rm V}=\omega^{-2}
 \left|\left\langle{\mit\Psi}_f
 \vert\sum_{j=1}^{N+1} \frac{\partial}{\partial r_j}\vert
 {\mit\Psi}_i\right\rangle\right|^2. \label{eq:SVe}
\end{equation}
In these equations $\omega$ is the incident photon energy
in Rydberg units, and $\mit\Psi_i$ and $\mit\Psi_f$ are the wave 
functions representing the initial and final states respectively. The 
boundary conditions satisfied by a bound state with negative energy
correspond to exponentially decaying partial waves in all `closed'
channels, whilst those satisfied by a free or continuum state
correspond to a plane wave in the direction of the ejected electron 
momentum $\underline{\hat{k}}$ and ingoing waves in all open channels.

Using the energy difference, $E_{ji}$, between the initial and final
states, the oscillator strength, $f_{ij}$, for the transition can be 
obtained from $S$ as

\begin{equation}
f_{ij} = {E_{ji}\over {3g_i}}S,
\end{equation}

\noindent
and the Einstein's A-coefficient, $A_{ji}$, as

\begin{equation}
A_{ji}(a.u.) = {1\over 2}\alpha^3{g_i\over g_j}E_{ji}^2f_{ij},
\end{equation}

\noindent
where $\alpha$ is the fine structure constant, and $g_i$, $g_j$ are 
the statistical weight factors of the initial and final states,
respectively. In terms of c.g.s. unit of time,
\begin{equation}
A_{ji}(s^{-1}) = {A_{ji}(a.u.)\over \tau_0},
\end{equation}

\noindent
where $\tau_0 = 2.4191^{-17}$s is the atomic unit of time.

\section{Computations}

The target wavefunctions of Fe VI were obtained by Chen and Pradhan [21]
from an atomic structure calculation using the Breit-Pauli version of 
the SUPERSTRUCTURE program [22], intended for electron collision
calculations with Fe~VI using the Breit-Pauli R-matrix method.
Present work employs their optimized target of 19 fine 
structure levels [21] corresponding to the 8-term $LS$ basis set of 
$3d^3 (^4F$, $^4P$, $^2G$, $^2P$, $^2D2$, $^2H$, $^2F$, $^2D1)$. The set of 
correlation configurations used were $3s^23p^63d^24s$, $3s^23p^63d^24d$, 
$3s3p^63d^4$, $3p^63d^5$, $3s^23p^43d^5$, and $3p^63d^44s$. The values
of the scaling parameter in the Thomas-Fermi potential for 
each orbital of the target ion are given in Ref. [21]. 
Table I lists the 19 fine structure energy levels of Fe~VI used in the 
eigenfunction expansion where the energies are the observed ones. 
Most bound levels in low ionization stages correspond to the level of 
excitation of the parent ion involving the first few excited states.
The criterion remains the accuracy of the 
target represetation that constitute the core ion states.  The (N+1) 
electron configurations, $\Phi_j$, which meet the orthogonality condition 
for the CC expansion (the second term of the wavefunction, Eq. (1)) 
are given below Table I. The same set of configurations is used
for all the states considered in this work.  STG1 of the BPRM codes 
computes the one- and two-electron radial 
integrals using the one-electron target orbitals generated by 
SUPERSTRUCTURE. The number of continuum basis functions is 12.

The present calculations are concerned with all possible bound levels 
with $n\leq 10, \ \ell \leq n-1$. These correspond to total (e + Fe~VI) 
symmetries ($SL\pi$) with (2$S$ + 1) = 1,3,5 and $L$ = 0 - 10 (even and 
odd parities).  The intermediate coupling calculations 
are carried out on recoupling these $LS$ symmetries in a pair-coupling 
representation, Eq. 6, in stage RECUPD.  The computer memory requirement
for this stage is the maximum, since it carries out angular algebra of
dipole matrix elements of a large number of fine structure levels.
The (e + Fe~VI) Hamiltonian is diagonalized for each 
resulting $J\pi$ in STGH. The negative eigenvalues of the Hamiltonian 
correspond to the bound levels of Fe~V, that are found according to 
the procedure described below.

\subsection{Calculation of bound levels}

The eigenenergies of the Hamiltonian for each $J\pi$ are determined 
with a numerical search on an effective quantum number mesh, with an 
interval $\Delta \nu$, using the code STGB. In the relativistic case, the
number of Rydberg series of levels increases considerably from those 
in $LS$ coupling due to splitting of the target states into their fine
structure components. This results in a large number of fine structure 
levels in comparatively narrow energy bands. A mesh with 
$\Delta (\nu) = 0.01$ is usually adequate to scan for $LS$ term 
energies; however, it is found to be of insufficient resolution for
fine structure energy levels. The mesh needs to be finer by an order
of magnitude, i.e., $\Delta (\nu) = 0.001$, so as not to
miss out any significant number of bound levels. This considerably
increases the computational requirements for the intermediate coupling
calculations of bound levels over the LS coupling case by
orders of magnitude. The calculations take up to several CPU hours per 
$J\pi$ in order to determine the corresponding eigenvalues. 
All bound levels of total $J \leq$ 8, of both parities, are 
considered. However, a further search with an even finer $\Delta \nu$
reveals that a few levels are still missing for some $J\pi$ symmetries.

\subsection{Procedure for level identification}

The energy levels in the BPRM approximation (from STGB) are identified
by $J\pi$ alone. This is obviously insufficient information to 
identify all associated quantum numbers of a level from among a large 
set levels for each $J\pi$, typically a few hundred for Fe~V. A sample 
set of energy levels for $J = 2$, even parity, obtained from the BPRM 
calculations is presented in Table II. The table shows energies 
and effective quantum number $\nu_g$, as calculated relative to the 
ground level $(3d^{3~4}F_{3/2})$ of the core ion Fe~VI.
The complexity of the calculations, and that of level identification,
may be gauged from the fact that 30 of these levels have nearly the same
$\nu_g$. Further, the $\nu_g$ do not in general correspond to the 
actual effective quantum number of the Fe~V level since it may belong 
to an excited parent level, and not the ground level, of Fe~VI.

A scheme has been developed to identify the levels with complete 
spectroscopic information consisting of 
\begin{equation}
(C_t \ S_t \ L_t \ J_t~\pi_t \ell \ [K] {\rm s})\ \ J \ \pi, 
\end{equation}
and also to designate the levels with a possible $SL\pi$ symmetry. 
The designation of the $SL\pi$, from the 
identifications denoted above, is generally ambiguous since the
collision channels are all in intermediate coupling. However,
in most cases we are able to carry through the identification procedure
to the $LS$ term designation. An advantage of identification is that 
it greatly facilitates the completeness check for all possible LS terms 
and locate any missing levels. A computer code PRCBPID has been developed 
to identify all the quantum numbers relevant to the $J\pi$ and the $LS$
term assignments. Identification is carried out for all the levels 
belonging to a $J\pi$ symmetry at a time.

The components of the total wavefunction of a given fine structure 
energy level span all closed "collision" channels 
$(C_tS_tL_t(J_t)\pi_t)\epsilon l$. Each channel contains the information
of the relevant core and the outer electron angular momentum. 
The ``channel weights", mentioned earlier, determine the magnitude of 
the wavefunction in the outer R-matrix region of each channel 
evaluated in STGB. A bound level may be readily assigned to the 
quantum numbers of a given channel provided the corresponding channel 
weight (in percentage terms) dominates the other channels. 

The number of channels can be large especially for complex ions. 
For Fe V, for example, each level with $J~>~2$ corresponds to several
hundred channels.  As the first step in the level identification 
scheme we isolate the two most dominant channels by comparing all
channel percentage weights. The reason is that the largest channel 
percentage weight may not uniquely
determine the identifications since the channel weights are 
evaluated from the outer region contributions ($r > a$); the inner region 
contributions are unknown. Also, many levels are often heavily mixed and 
no assignment for the dominant channel may be made.

The program, PRCBPID, sorts out the duplicate identifications in all 
the levels of the $J\pi$ symmetry. Two levels with the same 
configuration and set of quantum numbers can actually be two 
independent levels due to outer electron spin addition/ subtraction to/from 
the parent spin angular momentum, i.e. $S_t\pm s = S$. The identical pair
of levels are tagged with positive and negative signs indicating higher
and lower multiplicity respectively. The lower energies are normally 
assigned with the higher spin multiplicity. However, the energies and 
effective quantum numbers ($\nu$) of levels of higher and lower spin 
multiplicity  can be very close to each other, in which case the spin 
multiplicity assignment may be uncertain.

One important identification criterion is the analysis of the quantum 
defect, $\mu$, or the effective quantum number, $\nu$, of the outer or 
the valence electron. The principle quantum number, $n$, of the outer 
electron of a level is determined from its $\nu$, and a Rydberg series 
of levels can be identified from the effective quantum number. Hence,
in the identification procedure, $\nu$ of the lowest member (level with 
the lowest principal quantum number of the valence electron) of 
a Rydberg series is determined from quantum defect analysis of all 
the computed levels for each partial wave $l$. The lowest partial wave 
has the highest quantum defect. A check is maintained to differentiate 
the quantum defect of a $'s'$ electron with that of an equivalent electron 
state which has typically a large value in the close coupling calculations. 
The principle quantum number, $n$, of the lowest member of the series is 
determined from the orbital angular momentum of the outer electron and 
the target or core configuration. Once $\nu$ and $\mu = n-\nu$ of the 
lowest member are known, the $n$-values of all levels can be assigned 
for each paritial wave, $l$. The relevant Rydberg series of levels is 
also identified from the levels that have the same symmetry, $J\pi$, 
core configuration, $C_t \ S_t \ L_t \pi_t$ and outer electron orbital 
angular momentum $l$, but different $\nu$ that 
differs between successive levels by $\sim$ 1.
While the $\nu (n \ \ell)$ are more accurate for the higher members of the 
series, they are more approximate for the lowest ones. 
The quantum defect of a given partial wave $\ell$ also varies slightly 
with different parent core levels and final $SLJ$ symmetries. 

Of the two most dominant channels the proper one for each bound level 
is determined based on several criteria. 
There are cases when more than two levels are found to have identical 
identifications. These levels are checked individually for proper 
identification.  Often a swap of identifications is needed between the two 
sets of dominant channels since the second dominating channel is more
likely to be associated with the given level, consistent with all other
criteria. In some cases the most dominant channel (largest percentage weight 
in the outer region) may correspond to comparatively larger $\nu$ 
for the partial wave $\ell$, than to a reasonable $\nu$ for the 
second channel, indicating that the identification should correspond 
to the second channel.

In a few cases a level is found not to correspond to any 
of the two dominant channels, predetermined from the channel weights. 
At the same time often a level is found to be missing in the same energy
range.
In such case the level is assigned to a channel of lower percentage 
weight that has a reasonable core configuration 
and term, $nl$ quantum numbers for the outer electron and effective 
quantum number that match the missing level. 

There are a number of levels belonging to equivalent-electron 
configurations and require different identification criteria from 
those of the Rydberg states. These levels usually have: (i) a number 
of approximately equal channel weights, and (ii) quantum defects that 
are larger than that of the lowest partial wave, or an inconsistent 
$\nu$ that does not match with any reasonable $n$. Once these levels 
are singled out, they are identified with the possible configurations 
of the core level, augmented by one electron in the existing orbital 
sub-shell.  These low-lying levels are often assigned to those 
identified from the small experimentally available set of observed 
levels. The levels that can not be identified in the above procedure,
such as by swapping of channels, or maching to a missing level, are
assumed to belong to mixed states. These are not analysed futher by 
quantum defects.

Two additional (and related) problems, as mentioned above, are 
addressed in the identification work: (A) standard $LS$ coupling 
designation, $SL\pi$, and (B) the completeness check for the set of all 
fine structure components within an $LS$ multiplet. 
Identification according to collision channel quantum numbers is not
quite sufficient to establish a direct correspondence with the standard
spectroscopic notation employed in atomic structure calculations, or in
the compiled databases such as those by the U.S. National Institute for
Standards and Technology (NIST).  

The possible set of $SL\pi$s of a level is obtained from the target 
term, $S_tL_t\pi_t$, and the valence electron angular momentum, $l$, 
at the first occurance of the level in the set. The total 
spin multiplicity of the level is defined according to the energy level
position as discussed above.  For example, the core $3d^3(^4F^e)$ 
combining with a $4d$ electron forms the terms $^5(P, D, F, G, H)^e$ 
and $^3(P, D, F, G, H)^e$ (Table IV) where the quintet for each $L$ 
should be lower than the triplet. 

To each $LS$ symmetry, $SL\pi$, of the set belongs a set of predetermined 
$J$-levels. The set of total J-values of same spin multiplicity is 
then calculated from all possible $LS$ terms, equal to $|{\bf L+S}|$. 
The program sorts out all calculated fine stucture levels with the same 
configuration, but with different sets of $J_t$ and $J$, e.g.  
$(C_t \ S_t \ L_t \ J_t~\pi_t n\ell )\ \ J \ \pi$ (including the sign for
the upper or lower spin multiplicity), compares them with the 
predetermined set, and groups them
together. Thus a correspondence is made between the set of $SL\pi$ 
and the calculated fine structure levels of same configuration. 

In addition to the correspondence between the two sets, the program 
PRCBPID also calculates the possible set of $SL\pi$'s for each 
single J-level in above group. In the set of $SL\pi$s, the total 
spin is fixed while the angular momentum, $L$, varies. In the above 
example for the quintets,  $^5(P, D, F, G, H)^e$, each $J$=1 
level is assigned to a possible set of terms, $^5(P, D, F)$ 
(Table IV). However, these levels can be futher 
identified uniquely following Hund's rule that the term with the 
larger angular momentum, $L$, is the lower one, i.e., 
the first or the lowest $J$=1 level should correspond to $^5F$, the 
second one to $^5D$ and the last one to $^5P$.

The completeness of sets of fine structure levels with respect to the 
$LS$ terms are checked. As mentioned above, PRCBPID determines the 
possible sets of $SL\pi$ from the target term and valence electron 
angular momentum of a level at its first occurance and calculates the
total J-values of the set of $LS$ terms. The number of these $J$-values,
$Nlv$, is compared with that of calculated levels, $Ncal$ to check the 
completeness. For example, for the above case of $^5(P, D, F, G, H)^e$ 
in Table IV discussed above, $^5P$ can have $J$ = 1,2,3, $^5D$ can have
$J$ = 0,1,2,3,4, and so on, giving a total of 23 fine structure levels 
for this set of $LS$ terms. The one $J$ = 0 level belongs to $^5D$, the 
three $J$ = 1 levels belong to $^5(P, D, F)$, and so on. All 23 levels
of this set are found in the computed levels (Table IV), thus making 
the computed set complete. This procedure, in addition to finding the 
link between the two diiferent coupling schemes, enables an independent 
counting of the number of levels obtained, and ascertains missing 
or mis-identified levels.

\subsection{Transition probabilities}

The oscillator strengths (f-values) and transition probabilites
(A-values) for bound-bound fine structure level transitions in Fe V
are calculated for levels up to $J \leq 8$. Computations are carried out 
using STGBB of the BPRM codes. 

The f-values are initially calculated by the program STGBB with level 
designations given by $J\pi$ only. However, the transitions may be 
fully described following the level
identifications as described in the previous section. 
Work is in progress to identify all the transitions with proper
quantum numbers, configurations and possible $SL\pi$'s. 

A subset of the large number of transitions has been processed with
complete identifications. Among these transitions are those that 
correspond to the experimentally observed levels [23]. 
As these levels have been identified, their oscillator
strenghts could be sorted out from the file of f-values. Another
subsidiary code, PRCBPRAD, is developed to reprocess the transition 
probabilities where the calculated transition energies are replaced
by the observed ones for improved accuracy.

The computation time required for the BPRM calculations was orders
of magnitude longer compared to oscillator strengths calculations in LS 
coupling, as carried out under the OP for example. The time excludes that
needed for creating the necessary bound state wavefunctions and calculating
dipole matrix elements using the R-matrix package of codes. 
Computations are carried out for one or a few pairs of symmetries
at a time requiring several hours of CPU time on the Cray T94. The memory
requirement was over 30 MWords.

\section{Results and discussion}

Theoretical spectroscopic data are calculated on a large-scale with 
relativistic fine structure included in an {\it ab initio} manner, 
and ensuring completeness in terms of obtaining nearly all possible 
energy levels and transition probabilities for Fe~V for the total 
angular symmetries considered. The results are described below.

\subsection{Energy Levels}

We have calculated 3,865 fine structure bound levels, with 0 $\leq J
\leq$ 8, for Fe V. Following level identification, as explained in the 
previous section, the energy levels are arranged according
to ascending order in energy. 

The present energies are compared with the relatively small set of 
experimentally observed levels compiled by NIST [23] in Table III. 
All 179 observed levels are obtained and identified. Asterisks 
attached to levels in Table III indicate an incomplete set of observed 
levels corresponding to the $LS$ term. Often in experimental measurements
the weak lines are not observed. The theoretical datasets on the other
hand are usually complete.

We find some discrepancies regarding the identification of 
a couple of levels in the NIST tabulation. The $J=2$ level at 2.9395 Ry 
identified in the NIST table as $3d^3(^4P)4p(^5S^o)_2$, from the maximum 
leading percentage, may have been misidentified. Present analysis for 
the completeness of a set of fine structure levels belonging to a 
term indicates it as an extra level for the given configuration and 
that the possible $LS$ terms for this level are $3d^3(^2D2)4p(^3PDF^o)$, 
possibly $^3F^o$. Similarly the NIST identification for the $J$=3 level at 
2.8968 Ry is $3d^3(^2P)4p(^3D^o)_3$, from the maximum leading percentage. 
Present calculations however assign the level to possible LS terms,
$3d^3(^2D2)4p(^3DF^o)$, and most likely to $^3D^o$. 

In the computed set of fine structure levels the observed levels are 
usually the ones with the lowest energy in each subset of $J\pi$. The
lowest calculated levels are the 34 levels of the ground configuration 
$3d^4$ of Fe~V, in 
agreement with the observed ones. The agreement between the observed and
calculated energies for these levels is within 1\%. The 
calculated energies agree to about 1\% with the measured ones for most 
of the observed levels. Although the energies are exoected to be highly 
accurate, but the uncertainty in the calculations is not comparable to
that in spectroscopic observations (of the order of few wavenumbers).

Employing the completeness procedure the computed fine structure 
levels are tabulated, according to the two sets of cross-correlating 
quantum numbers: one according to the collision channels identified as
$(C_tS_t \ L_t \ J_t \pi_t~n\ell \ [K] {\rm s})\ \ J \ \pi$, and the 
other according to the complete set of J-values for each multiplicity 
$(2S+1)$, $L$ and $\pi$. A subset of the complete table of fine 
structure levels is presented in Table IV. (The complete table will 
be available electronically). Each set of levels is grouped by the 
possible set of $LS$ terms followed by the levels of same configuration,
core term, total spin multiplicity and parity, and with different 
$J$-values. The header for each group contains the total
number of possible $J$-levels, $Nlv$, total spin multiplicity, parity, 
and all possible $L$ values formed from the core and the outer 
electron. The possible $J$-values for each $SL\pi$ are given within 
parentheses next to each $L$ value. 

The two sets of quantum numbers are compared. The levels that may be 
missing or mis-identified are thereby checked out. The
number of computed levels, $Ncal$, is compared with that expected
from angular and spin couplings, $Nlv$. For most of the
configurations the set of levels is complete except for the high lying
ones. The comparison detects missing levels. An example is shown in the
the set of $3d^32(^2D)5d^3(S,P,D,F,G)^e$ in Table~IV where one level
with $J^e$ = 4 is missing.

In Table IV, the effective quantum number $\nu$ is specified alongwith
other quantum numbers for each level. The
consistency in $\nu = {z\over \sqrt(E-E_t)}$, where $E_t$ is the 
corresponding target energy, for each set of levels may be noted. 

The possible $SL\pi$s for each level are given in the last column.
The levels with a single possible term only are uniquely defined.
However, those with two or multiple term assignments can be defined 
uniquely applying Hund's rule that the higher $L$ corresponds to the 
lower energy of same $J\pi$ as explained in the previous section (we 
note that Hund's rule may not always apply in cases of strong CI).

There are 112 levels of odd parity that we could not properly identify.
Some of these levels are given in Table IV. These 
levels could be equivalent electron levels of configuration, 
$3p^5(^2P^o)3d^5$. The 16 LS terms of $3d^5$, which are $^2D1$,
$^2P3$, $^2D3$, $^2F3$, $^2G3$, $^2H3$, $^4P5$, $^4F5$, $^2S5$, $^2D5$,
$^2F5$, $^2G5$, $^2I5$, $^4D5$, $^4G5$, and $^6S5$, in
combination with the parent core $^2P^o$, form 88 LS terms with 31 singlets, 
43 triplets, 13 quintets and 1 septet. The number of fine structure 
levels from these terms exceed the 112 computed levels that have not 
been identified.

This new procedure of cross-correlation between two coupling schemes
thus provides a powerful check on the completeness and level
identification, and is expected to be of use in further BPRM work 
on complex atomic systems.

\subsection{Transition Probabilities}

The oscillator strengths (f-values) and transition probabilites
(A-values) for fine structure level transitions in Fe V are obtained
for $J \leq 8$. The allowed $\Delta J=0,\pm 1$ transitions include 
both the dipole allowed ($\Delta S=0,\pm1$) and the intercombination 
($\Delta S \neq$ 0) transitions. The total number of computed 
transition probabilities is well over a million, 
approximately $1.46\times 10^6$. For most allowed pairs of $J\pi$ 
symmetries, there are about $10^3~-~10^5$ transitions.  

As explained in the previous section, a subset of the encoded transitions 
have been processed to present them with proper identifications. These
correspond to the levels that have been observed. A sample of these
is presented in Table~V. In all of the f-values presented the calculated 
transition energy has been replaced by the observed one, using
the BPRM line strengths ($S$) which are energy independent.
Since measured 
energies in general have smaller uncertainties than the calculated 
ones, this replacement improves the accuracy of the oscillator 
strengths. The transitions among the 179 observed levels correspond to 
3727 oscillator strengths. (The complete set of transition probabilities 
will be available electronically.) 

The f-values in Table~V have been reordered to group the transitions of
the same multiplet together. This enables a check on the completeness of
the set of transitions. As this table corresponds to transitions among
observed levels only, the completeness depends on the set of observed
levels belonging to the LS terms. For the dipole allowed transitions,
the LS multiplets are also given at the end of $jj'$ transitions.

To our knowledge, no measured f-values for Fe V are available for 
comparison.  Current NIST compilation [1] contains no f-values 
for any allowed transition. On the other hand, Fe~V oscillator 
strengths for a large number of transitions were obtined in the 
close coupling approximation under the OP [24] and the IP [25]. 
Both of these datasets are non-relativistic  calculations in LS 
coupling and do not compute fine structure transitions. Fawcett 
has [26] carried out semi-empirical relativistic atomic structure 
calculations for fine structure transitions in Fe V. Comparison of 
the present $f$-values is made with the previous ones in Table VI, 
showing various degrees of agreement. Present values agree within 10\%
with those by Fawcett for a number of fine structure transitions of
multiplets, $3d^4(^5D)\rightarrow 3d^3(^4F)4p(^5D^o)$, and
$3d^4(^5D)\rightarrow 3d^3(^4P)4p(^5P^o)$, and the disagreement is large
with other as well as with those of $3d^4(^5D)\rightarrow 
3d^3(^4F)4p(^5F^o)$. The agreement of the present $LS$ multiplets 
with the others is good for transitions $3d^4(^5D)\rightarrow 
3d^3(^4F)4p(^5F^o,^5D^o,^5P^o)$. More 
detailed comparisons will be made at the completion of this work.

The procedure of substitution of experimental for calculated
energies provides an indication of uncertainties in the calculated 
$f$-values. The difference between the $f$-values obtained using 
the calculated transition energies and the observed ones is only a few
percents ($<$ 5\%) for most of the allowed transitions. The difference 
is usually larger for the intercombination transitions which have 
lower transition probabilities. In atomic structure calculations, 
it is possible to re-adjust eigenenergies of the Hamiltonian to match 
the observed ones and then use the wavefunctions to obtain the
transition probabilities. Such a  re-adjustment is not carried out 
in the BPRM calculations of bound states, which are entirely ab initio,
with the associated advantage of consistent uncertainties 
for most transitions considered.

To obtain an estimate of the accuracy of the wavefunctions employed 
in the length and the velocity formulations, we plot, for example, 
the $gf$-values for transitions $(J=1)^e-(J=2)^o$ and $(J=3)^e-(J=4)^o$
in Fig. 1. The top panel contains over 13,300 transitions between the
pair of symmetries $(J=1)^e~-~(J=2)^o$, and the bottom panel contains 
over 20,200 transitions
between the pair $(J=3)^e~-~(J=4)^o$. The plots show practically 
no dispersion for the strongest transitions with
$gf \approx 5 - 10$, and some dispersion  around 10-20\% for others 
with $gf < 3$. Up to $gf < 0.1$ the dispersion in length and velocity
remains around the 10-20\% level for most of the transitions, although
the number of outlying transitions increases with decreasing $gf$.
Given the large number of points in the figures, the 
relatively low dispersion of $gf_L$ and $gf_V$ indicates that the 
$f$-values ($gf$ divided by the statistical weight factor, $2J$+1) for 
most of the transitions with $gf\sim$1 should be within 20\% uncertainty. 
The $f_L$'s are usually more accurate than $f_V$'s since the asymptotic 
region wavefunctions are more accurately represented 
in the close coupling calculations using the R-matrix method.

In general the intercombination transitions are weaker than the dipole
allowed ones; the f-values can be orders of magnitude lower.
The BP Hamiltonian in the present work (Eq. 2) does not include the 
two-body spin-spin and spin-other-orbit terms of Breit interaction [22]. 
A discussion of these terms is given
by Mendoza \etal in a recent IP paper [27]. Their study on the 
intercombination transitions in C-like ions shows that
the effect of the two-body Breit terms, relative to the one-body
operators,  decreases with Z such that for Z = 26 the computed A-values
with and without the two-body Breit terms differ by less than 0.5 \%.
However, the differences towards the neutral end of the C-sequence is
up to about 20\%. It may therefore be expected that for Fe~V 
the weaker intercombination f-values may also be 
systematically affected to a similar extent (the uncertainties in the 
dipole allowed f-values should be much less). Further studies of the
Breit interaction in complex atoms are needed to ascertain this effect
more precisely. 

Several aspects of the present work are targets for future studies, 
such as atomic structure calculations to study the effect of 
configuration interaction and relativistic effects on different types 
of transitions, and a detailed
quantum defect analysis along interacting Rydberg series of levels in
intermediate coupling. These studies should provide information on the
accuracy of particular type of transitions and groups of levels, as well
as address general problems in the analysis of complex spectra.

\section{Conclusion}

The present work is the first study of large-scale transition
probabilities computed using the accurate BPRM method for a highly
complex ion. Some of the results obtained herein are expected to form 
the basis for future computational spectroscopy of heretofore 
intractable complex atomic systems using efficient collision theory 
methods. The computational procedures developed for such undertakings 
are described, and illustrative results are presented from the 
{\it ab initio} Breit-Pauli R-matrix calculations for Fe~V. Detailed 
analysis for the identification of over 3,800 fine structure levels of 
Fe V is carried out using a combination of methods that include 
quantum defect theory. Further work on the analysis of relativistic 
quantum defects in intermediate coupling is planned.

Following the completion of all computations and identifications, the
dataset of approximately 1.5 million oscillator strengths will be 
described in another publication with a view towards 
astrophysical and laboratory applications. In order to complete the
dataset for practical applications calculations are also in progress
for the forbidden electric quadrupole and magnetic dipole transition
probabilties using the atomic structure program SUPERSTRUCTURE. 

The newly acquired theoretical capability to  obtain an essentially
complete description of radiative transitions for an atomic system should
enable several new advances such as: (a) the synthesis of highly detailed
monochromatic opacity spectra [2], (b) the simulation 
of ``quasi-continuum" line spectra from iron ions [28], (c) high
resolution spectral diagnostics of iron in laboratory fusion and 
astrophysical sources, and (d) the analysis of experimentally measured 
spectra of complex iron ions.

\vskip 0.7in

\noindent
{\bf Acknowledgements}

\noindent
This work was supported partially by the U.S. National Science
Foundation (AST-9870089) and the NASA Astrophysical Theory Program. 
The computational work was carried out at the Ohio Supercomputer Center 
in Columbus Ohio.

\vskip 0.4in
\noindent
{\bf References \\}
\parindent=0.0in
\renewcommand{\baselinestretch}{1.0}
\def\amp{{\it Adv. At. Molec. Phys.}\ }
\def\apj{{\it Astrophys. J.}\ }
\def\apjs{{\it Astrophys. J. Suppl. Ser.}\ }
\def\apjl{{\it Astrophys. J. (Letters)}\ }
\def\aj{{\it Astron. J.}\ }
\def\aa{{\it Astron. Astrophys.}\ }
\def\aasup{{\it Astron. Astrophys. Suppl.}\ }
\def\adndt{{\it At. Data Nucl. Data Tables}\ }
\def\cpc{{\it Comput. Phys. Commun.}\ }
\def\jqsrt{{\it J. Quant. Spectrosc. Radiat. Transfer}\ }
\def\jpb{{\it Journal Of Physics B}\ }
\def\pasp{{\it Pub. Astron. Soc. Pacific}\ }
\def\mn{{\it MNRAS}\ }
\def\psc{{\it Physica Scripta}\ }
\def\pra{{\it Physical Review A}\ }
\def\prl{{\it Physical Review Letters}\ }
\def\zpds{{\it Z. Phys. D Suppl.}\ }

1. Fuhr, J.R., Martin, G.A., Wiese, W.L., J. Phys. Chem. Ref. Data {\bf
17}, Suppl No. 4 (1988) \\
2. Seaton, M.J., Yu, Y., Mihalas, D., Pradhan, A.K., MNRAS {\bf 266}. 
805 (1994). \\
3. Rogers F.J. and Iglesias C.A., {\it Science} {\bf 263}, 50 (1994)\\
4. Chayer, P., Fontaine, G., and Wesemael, F., Astrophys.J. {\bf 99},
189 (1995).\\
5. Becker, S.R. and Butler, K., \aa, {\bf 265}, 647 (1992)\\
6. Vennes, S., {\it Astrophysics in the Extreme
Ultraviolet} (Ed: Stuart Bowyer and Roger F. Malina), Kluwer, p. 185;
(1996); Pradhan, A.K., Ibid., p. 569\\
7. Seaton M.J. 1987, J.Phys.B {\bf 20}, 6363. \\ 
8. Hummer, D.G., Berrington, K.A., Eissner, W., Pradhan, A.K., Saraph,
H.E., Tully, J.A., Astron. Astrophys. {\bf 279}, 298 (1993)\\
9. Burke, P.G., Hibbert, A., and Robb, \jpb {\bf 4}, 153 (1971) \\
10. Berrington K.A., Burke P.G., Butler K., Seaton M.J., Storey P.J., Taylor
K.T. and Yu Yan, J.Phys.B {\bf 20} 6379 (1987).\\
11. Nahar S.N. and Pradhan, A.K., Astron. Astrophys. Suppl. Ser. 135, 
347 (1999). \\
12. Zhang, H.L., \pra {\bf 57}, 2640 (1998) \\
13. Zhang H.L., Nahar S.N., and Pradhan A.K., J. Phys. {\bf B 32}, 
1459 (1999).\\
14. Johnson W.R., Liu Z.W., and Sapirstein J., At. Data Nucl. Data 
{\bf 64}, 279 (1996). \\
15. Yan Z-C, Tambasco M., and Drake G.W.F., Phys. Rev. A {\bf 57}, 
1652 (1998) \\
16. Burke P.G. and Seaton M.J., \jpb {\bf 17}, L683 (1984) \\
17. Seaton M.J. J.Phys.B {\bf 18}, 2111 (1985) \\ 
18. Scott N.S., Taylor K.T., Comput. Phys. Commun. {\bf 25}, 347 (1982) \\
19. Berrington K.A., Eissner, W., Norrington P.H. Comput. Phys.
Commun. {\bf 92}, 290 (1995) \\
20. Burke, P.G. and Berrington, K.A., {\it Atomic and Molecular
Processes, an R-matrix Approach}, Institute of Physics Publishing,
Bristol (1993)\\
21. Chen, G.X. and Pradhan, A.K., \jpb {\bf 32}, 1809 (1999a); \aasup
{\bf 136}, 395 (1999b)\\
22. Eissner, W., Jones, M., Nussbaumer, H., Comput. Phys. Commun {\bf 8},
270 (1974); W. Eissner, J. Phys. IV (Paris) C1, 3 (1991), Eissner W. 
(in preparation, 1999) \\
23. Sugar, J. and Corliss, C., J. Phys. Chem. Ref. Data 14, Suppl. 
{\bf 2} (1985) \\
24. Butler, K. (unpublished); data are available through the OP database,
TOPbase (W. Cunto, C. Mendoza, F. Ochsenbein, C.J. Zeippen, Astron. 
Astrophys {\bf 275}, L5 (1993)) \\
25. Bautista M.A., A\&A Suppl. Ser. {\bf 119}, 105 (1996) \\
26. Fawcett, B.C. At. Data Nucl. Data Tables {\bf 41}, 181 (1989) \\
27. Mendoza C., Zeippen C.J. and Storey P.J., A\&A Suppl.Ser. 135, 159 
(1999) \\
28. Beirsdorfer, P. Lepson, J.K., Brown, G.V., Utter, S.B., Kahn, S.M., 
Liedahl, D.A., and Mauche, C.W., Astrophys. J. {\bf 519}, L185
(1999)\\

\vskip 3.0in

\noindent
Figure captions:

1. Comprarison of $gf_L$ versus $gf_V$ for bound-bound fine structure
level transitions in Fe V obtained in BPRM approximation.

\pagebreak

\begin{table}
\noindent{Table I. {\it The 19 fine strucuture levels of Fe IV in the 
close coupling eigenfunction expansion of Fe V. List of configurations,
$\Phi_j$, in the second sum of $\Psi$ is given below the table.} \\ }
\small
\begin{tabular}{lrrlrr}
\noalign{\smallskip}
\hline
\noalign{\smallskip}
Term & $J_t$ & $E_t$(Ryd) & Term & $J_t$ & $E_t$(Ryd) \\
\noalign{\smallskip}
\hline
\noalign{\smallskip}
$3d^3(^4F)$ & 3/2 & 0.0 &
$3d^3(^2P)$ & 1/2 & 0.241445 \\
$3d^3(^4F)$ & 5/2 & 0.004659 &
$3d^3(^2D2)$ & 5/2 & 0.259568 \\
$3d^3(^4F)$ & 7/2 & 0.010829 &
$3d^3(^2D2)$ & 3/2 & 0.260877 \\
$3d^3(^4F)$ & 9/2 & 0.018231 &
$3d^3(^2H)$ & 9/2 & 0.261755 \\
$3d^3(^4P)$ & 1/2 & 0.170756 &
$3d^3(^2H)$ & 11/2 & 0.266116 \\
$3d^3(^4P)$ & 3/2 & 0.172610 &
$3d^3(^2F)$ & 7/2 & 0.421163 \\
$3d^3(^4P)$ & 5/2 & 0.178707 &
$3d^3(^2F)$ & 5/2 & 0.424684 \\
$3d^3(^2G)$ & 7/2 & 0.187871 &
$3d^3(^2D1)$ & 5/2 & 0.653448 \\
$3d^3(^2G)$ & 9/2 & 0.194237 &
$3d^3(^2D1)$ & 3/2 & 0.656558 \\
$3d^3(^2P)$ & 3/2 & 0.238888 &
 & & \\
\noalign{\smallskip}
\hline
\noalign{\smallskip}
\multicolumn{6}{l}{$\Phi_j:$, $3s^23p^63d^4$, $3s^23p^63d^34s$, 
$3s^23p^63d^34p$, $3s^23p^63d^34d$, }\\
\multicolumn{6}{l}{$3s^23p^53d^44s$, $3s^23p^53d^44d$, $3s^23p^53d^34s$,
$3p^53d^54s4d$, } \\
\multicolumn{6}{l}{$3p^63d^54p$, $3s^23p^43d^6$, $3s^23p^43d^54p$, 
$3s3p^63d^34s4d$, } \\
\multicolumn{6}{l}{$3p^63d^34s4p^2$, $3s^23p^43d^44p^2$, $3s^23p^43d^44s^2$,
$3s3p^63d^44s$, } \\
\multicolumn{6}{l}{$3s^23p^53d^5$, $3p^63d^6$, $3p^63d^34s^24d$,
$3s3p^63d^44d$} \\
\end{tabular}
\end{table}

\pagebreak

\begin{table}
\noindent{Table II. {\it Sample set of calculated energy levels
(in $z^2$-scale) of $J$ = 2. } \\ }
\small
\begin{tabular}{rrrr}
\noalign{\smallskip}
\hline
\noalign{\smallskip}
\multicolumn{1}{c}{$E(Ry)$} & \multicolumn{1}{c}{$\nu_g$} & 
\multicolumn{1}{c}{$E(Ry)$} & \multicolumn{1}{c}{$\nu_g$} \\
\noalign{\smallskip}
\hline
\noalign{\smallskip}
 -1.492946E-01 & 2.58808 &
 -1.455232E-01 & 2.62140 \\
 -1.421287E-01 & 2.65252 &
 -1.383809E-01 & 2.68820 \\
 -1.381112E-01 & 2.69083 &
 -1.373195E-01 & 2.69857 \\
 -1.355085E-01 & 2.71654 &
 -1.307289E-01 & 2.76576 \\
 -1.202324E-01 & 2.88396 &
 -1.185430E-01 & 2.90444 \\
 -9.421634E-02 & 3.25789 &
 -9.408765E-02 & 3.26012 \\
 -9.385297E-02 & 3.26419 &
 -9.375868E-02 & 3.26584 \\
 -9.098725E-02 & 3.31520 &
 -9.071538E-02 & 3.32016 \\
 -8.971136E-02 & 3.33869 &
 -8.651911E-02 & 3.39973 \\
 -8.627750E-02 & 3.40448 &
 -8.536568E-02 & 3.42262 \\
 -8.458038E-02 & 3.43847 &
 -8.448129E-02 & 3.44049 \\
 -8.420175E-02 & 3.44619 &
 -8.371297E-02 & 3.45624 \\
 -8.338890E-02 & 3.46295 &
 -8.317358E-02 & 3.46743 \\
 -8.306503E-02 & 3.46969 &
 -8.225844E-02 & 3.48666 \\
 -8.180410E-02 & 3.49633 &
 -8.135593E-02 & 3.50595 \\
 -7.919621E-02 & 3.55343 &
 -7.827940E-02 & 3.57418 \\
 -7.815480E-02 & 3.57703 &
 -7.770277E-02 & 3.58742 \\
 -7.708061E-02 & 3.60186 &
 -7.691196E-02 & 3.60581 \\
 -7.594121E-02 & 3.62879 &
 -7.370981E-02 & 3.68330 \\
 -7.315088E-02 & 3.69735 &
 -7.138001E-02 & 3.74293 \\
 -7.028267E-02 & 3.77204 &
 -6.977748E-02 & 3.78567 \\
 -6.955230E-02 & 3.79179 &
 -6.845060E-02 & 3.82218 \\
 -6.765945E-02 & 3.84446 &
 -6.721368E-02 & 3.85719 \\
 -6.518968E-02 & 3.91661 &
 -6.265635E-02 & 3.99501 \\
 -6.190890E-02 & 4.01905 &
 -6.111942E-02 & 4.04492 \\
 -5.831623E-02 & 4.14100 &
 & \\
\hline
\end{tabular}
\end{table}

\clearpage

\pagebreak

\begin{table}
\noindent{Table III. {\it Comparison of calculated and observed energy
levels of Fe V. } \\ }
\scriptsize
\begin{tabular}{lllrr}
\noalign{\smallskip}
\hline
\noalign{\smallskip}
Configuration & Term & $J$ & E(cal) & E(expt) \\
\noalign{\smallskip}
\hline
\noalign{\smallskip}
$3d^4 $        &     $^5D$ &    4   &       5.5493  &    5.5015 \\
$3d^4 $        &     $^5D$ &    3   &       5.5542  &    5.5058 \\
$3d^4 $        &     $^5D$ &    2   &       5.5580  &    5.5094 \\
$3d^4 $        &     $^5D$ &    1   &       5.5607  &    5.5119 \\
$3d^4 $        &     $^5D$ &    0   &       5.5621 &    5.5132 \\
$3d^42$        &     $^3P$ &    2   &       5.3247  &    5.2720 \\
$3d^42$        &     $^3P$ &    1   &       5.3389  &    5.2856 \\
\hline
\end{tabular}
\end{table}

\begin{table}
\noindent{Table III. {\it continues. } \\ }
\scriptsize
\begin{tabular}{lllrr}
\noalign{\smallskip}
\hline
\noalign{\smallskip}
Configuration & Term & $J$ & E(cal) & E(expt) \\
\noalign{\smallskip}
\hline
\noalign{\smallskip}
$3d^3(^2H)4s $ &     $^3H$ &    4   &       3.4398  &    3.5377 \\
$3d^3(^2P)4s $ &     $^1P$ &    1   &       3.4015  &    3.5131 \\
$3d^3(^2D2)4s$ &     $^1D$ &    2   &       3.3877  &    3.5027 \\
$3d^3(^2H)4s $ &     $^1H$ &    5   &       3.3894  &    3.4965 \\
$3d^3(^2F)4s $ &     $^3F$ &    4   &       3.2711  &    3.3841 \\
$3d^3(^2F)4s $ &     $^3F$ &    3   &       3.2695  &    3.3822 \\
$3d^3(^2F)4s $ &     $^3F$ &    2   &       3.2682  &    3.3806 \\
\hline
\end{tabular}
\end{table}

\begin{table}
\noindent{Table III. {\it continues. } \\ }
\scriptsize
\begin{tabular}{lllrr}
\noalign{\smallskip}
\hline
\noalign{\smallskip}
Configuration & Term & $J$ & E(cal) & E(expt) \\
\noalign{\smallskip}
\hline
\noalign{\smallskip}
$3d^3(^4P)4p $ &     $^5S^o~!$ &2   &       2.9231 &     2.9395 \\ 
$3d^3(^2G)4p $ &     $^1F^o$ &  3   &       2.9067 &     2.9382 \\ 
$3d^3(^4P)4p $ &     $^5S^o$ &  2   &       2.8995 &     2.9379 \\ 
$3d^3(^2P)4p $ &     $^3P^o$ &  2   &       2.8605 &     2.8668 \\ 
$3d^3(^2G)4p $ &     $^1H^o$ &  5   &       2.9006 &     2.9354 \\ 
$3d^3(^2P)4p $ &     $^3D^o$ &  3   &       2.8895 &     2.9117 \\ 
$3d^3(^2P)4p $ &     $^3D^o$ &  2   &       2.8944 &     2.9169 \\
\hline
\end{tabular}
\end{table}

\clearpage

\begin{table}
\noindent{Table IV. {\it Sample table of calculated and identified fine 
strucuture energy levels of Fe V. Nlv=total number of levels expected
for the possible $LS$ terms (specified as $2S+1$, $\pi$, and set of $L$ 
with $J$-values within paratheses), formed from the target 
term and $l$ of the outer electron, and Ncal = number of calculated
levels. $SL\pi$ in last column=possible $LS$ terms for each level. 
 } \\ }
\scriptsize
\begin{tabular}{lcrrrcrl}
\noalign{\smallskip}
\hline
\noalign{\smallskip}
$C_t$ & $S_tL_t\pi_t$ & $J_t$ & $nl$ & $J$ & E(cal) & $\nu$ & $SL\pi$ \\
 \noalign{\smallskip}
 \hline
 \noalign{\smallskip}
\multicolumn{1}{l}{Nlv=~ 5,~ 5,e:} &
 \multicolumn{7}{l}{ F ( 5 4 3 2 1 )
             } \\
 \noalign{\smallskip}
 3d3      &(4Fe)& 3/2& 4s&  1& -3.73515E+00 &  2.59 &  5~ F e       \\
 3d3      &(4Fe)& 5/2& 4s&  2& -3.73238E+00 &  2.59 &  5~ F e       \\
 3d3      &(4Fe)& 5/2& 4s&  3& -3.72820E+00 &  2.59 &  5~ F e       \\
 3d3      &(4Fe)& 7/2& 4s&  4& -3.72275E+00 &  2.59 &  5~ F e       \\
 3d3      &(4Fe)& 9/2& 4s&  5& -3.71610E+00 &  2.59 &  5~ F e       \\
 \noalign{\smallskip}
 \multicolumn{8}{l}{Ncal=~ 5~ : set complete
                                                                       }
\\
\hline
\end{tabular}
\end{table}

\begin{table}
\noindent{Table IV. {\it continues. } \\ }
\scriptsize
\begin{tabular}{lcrrrcrl}
\noalign{\smallskip}
\hline
\noalign{\smallskip}
$C_t$ & $S_tL_t\pi_t$ & $J_t$ & $nl$ & $J$ & E(cal) & $\nu$ & $SL\pi$
\\
 \noalign{\smallskip}
 \hline
 \noalign{\smallskip}
 \multicolumn{1}{l}{Nlv=~13,~ 5,e:} &
 \multicolumn{7}{l}{ P ( 3 2 1 ) D ( 4 3 2 1 0 ) F ( 5 4 3 2 1 )
             } \\
 \noalign{\smallskip}
 3d3      &(4Pe)& 1/2& 4d&  3& -2.13511E+00 &  3.29 &  5~ PDF e     \\
 3d3      &(4Pe)& 5/2& 4d&  1& -2.13294E+00 &  3.29 &  5~ PDF e     \\
 3d3      &(4Pe)& 1/2& 4d&  2& -2.11203E+00 &  3.31 &  5~ PDF e     \\
 3d3      &(4Pe)& 5/2& 4d&  4& -2.09728E+00 &  3.31 &  5~ DF e      \\
 3d3      &(4Pe)& 3/2& 4d&  0& -2.08337E+00 &  3.33 &  5~ D e       \\
 \noalign{\smallskip}
 \multicolumn{8}{l}{Ncal=~13~ : set complete }
\\
\hline
\end{tabular}
\end{table}

\begin{table}
\noindent{Table IV. {\it continues. } \\ }
\scriptsize
\begin{tabular}{lcrrrcrl}
\noalign{\smallskip}
\hline
\noalign{\smallskip}
$C_t$ & $S_tL_t\pi_t$ & $J_t$ & $nl$ & $J$ & E(cal) & $\nu$ & $SL\pi$
\\
 \noalign{\smallskip}
 \hline
 \noalign{\smallskip}
 \multicolumn{1}{l}{Nlv=~13,~ 3,e:} &
 \multicolumn{7}{l}{ S ( 1 ) P ( 2 1 0 ) D ( 3 2 1 ) F ( 4 3 2 ) G ( 5 4 3 )
             } \\
 \noalign{\smallskip}
 3d3 2    &(2De) &5/2 &5d  &1     & -1.05446E+00& 4.36 &  3~SPD e \\
 3d3 2    &(2De) &5/2 &5d  &5     & -1.04247E+00& 4.38 &  3~G e \\
 3d3 2    &(2De) &3/2 &5d  &3     & -1.04179E+00& 4.38 &  3~DFG e \\
 3d3 2    &(2De) &3/2 &5d  &1     & -1.03332E+00& 4.40 &  3~SPD e \\
 3d3 2    &(2De) &5/2 &5d  &4     & -1.03014E+00& 4.40 &  3~FG e \\
 3d3 2    &(2De) &5/2 &5d  &2     & -1.02889E+00& 4.40 &  3~PDF e \\
\hline
\end{tabular}
\end{table}

\clearpage

\begin{table}
\noindent{Table V. {\it Transition probabilties of Fe V among observed
fine structure levels. } \\ }
\scriptsize
\begin{tabular}{llllrrcccc}
\noalign{\smallskip}
\hline
\noalign{\smallskip}
\multicolumn{1}{c}{$C_i$} & \multicolumn{1}{c}{$C_f$} & $S_iL_i\pi_i$ & 
$S_fL_f\pi_f$ & $g_i$ & $g_f$ & $E_i$ & $E_f$ & $f$ & $A_{fi}$ \\
 & & & & & & (Ry) & (Ry) & & ($s^{-1}$) \\
 \noalign{\smallskip}
 \hline
 \noalign{\smallskip}
 $3d^4         $ & $-~3d^3(4F)4p   $ & $^5D^e$ & $^5F^o$ &  1 &  3 &   5.5132 &   3.1644 &   2.154E-01 &   3.18E+09 \\
 $3d^4         $ & $-~3d^3(4F)4p   $ & $^5D^e$ & $^5F^o$ &  3 &  3 &   5.5119 &   3.1644 &   3.790E-04 &   1.68E+07 \\
 $3d^4         $ & $-~3d^3(4F)4p   $ & $^5D^e$ & $^5F^o$ &  3 &  5 &   5.5119 &   3.1496 &   1.358E-03 &   3.65E+07 \\
 $3d^4         $ & $-~3d^3(4F)4p   $ & $^5D^e$ & $^5F^o$ &  5 &  3 &   5.5094 &   3.1644 &   4.617E-02 &   3.40E+09 \\
 $3d^4         $ & $-~3d^3(4F)4p   $ & $^5D^e$ & $^5F^o$ &  5 &  5 &   5.5094 &   3.1496 &   5.967E-02 &   2.67E+09 \\
 $3d^4         $ & $-~3d^3(4F)4p   $ & $^5D^e$ & $^5F^o$ &  5 &  7 &   5.5094 &   3.1443 &   1.462E-02 &   4.69E+08 \\
\hline
\end{tabular}
\end{table}


\begin{table}
\noindent{Table V. {\it continues. } \\ }
\scriptsize
\begin{tabular}{llllrrcccc}
\noalign{\smallskip}
\hline
\noalign{\smallskip}
\multicolumn{1}{c}{$C_i$} & \multicolumn{1}{c}{$C_f$} & $S_iL_i\pi_i$ & 
$S_fL_f\pi_f$ & $g_i$ & $g_f$ & $E_i$ & $E_f$ & $f$ & $A_{fi}$ \\
 & & & & & & (Ry) & (Ry) & & ($s^{-1}$) \\
 \noalign{\smallskip}
 \hline
 $3d^4         $ & $-~3d^3(2P)4p   $ & $^5D^e$ & $^3P^o$ &  3 &  5 &   5.5119 &   2.8668 &   2.963E-05 &   9.99E+05 \\
 $3d^4         $ & $-~3d^3(2P)4p   $ & $^5D^e$ & $^3P^o$ &  5 &  3 &   5.5094 &   2.9439 &   6.224E-04 &   5.48E+07 \\
 $3d^4         $ & $-~3d^3(2P)4p   $ & $^5D^e$ & $^3P^o$ &  5 &  5 &   5.5094 &   2.8668 &   2.903E-05 &   1.63E+06 \\
 $3d^4         $ & $-~3d^3(2P)4p   $ & $^5D^e$ & $^3P^o$ &  7 &  5 &   5.5058 &   2.8668 &   3.718E-05 &   2.91E+06 \\
 & & & & & & & & & \\
 $3d^4         $ & $-~3d^3(2D2)4p  $ & $^5D^e$ & $^1P^o$ &  1 &  3 &   5.5132 &   2.9073 &   1.306E-08 &   2.38E+02 \\
\hline
\end{tabular}
\end{table}

\pagebreak

\begin{table}
\noindent{Table VI. {\it Comparison of present $f$-values with the earlier
ones. } \\ }
\scriptsize
\begin{tabular}{llllrrrrll}
\noalign{\smallskip}
\hline
\noalign{\smallskip}
\multicolumn{1}{c}{$C_i$} &\multicolumn{1}{c}{$C_j$} & $S_iL_i\pi_i$ &
$S_jL_j\pi_j$ & $2J_i+1$ & $I_i$ & $2J_j+1$ & $I_j$ & $f_{ij}(P)$ &
$f_{ij}(others)$ \\
 \noalign{\smallskip}
\hline
 \noalign{\smallskip}
 3d4   &     -3d3(4F)4p & 5D0 &5F1 & 1 &1& 3 &1& 0.2154&0.163$^a$\\
 3d4   &     -3d3(4F)4p & 5D0 &5F1 & 3 &1& 3 &1& 3.790E-04&  \\
 3d4   &     -3d3(4F)4p & 5D0 &5F1 & 3 &1& 5 &3& 0.00136&  \\
 3d4   &     -3d3(4F)4p & 5D0 &5F1 & 5 &1& 3 &1& 0.04617&0.0126$^a$\\
 3d4   &     -3d3(4F)4p & 5D0 &5F1 & 5 &1& 5 &3& 0.05967&0.0596$^a$\\
 3d4   &     -3d3(4F)4p & 5D0 &5F1 & 5 &1& 7 &3& 0.01462&0.0138$^a$\\
 3d4   &     -3d3(4F)4p & 5D0 &5F1 & 7 &1& 5 &3& 0.006895&0.0274$^a$\\
\end{tabular}
\end{table}

\end{document}